\documentclass[epj]{svjour}
\usepackage{graphics}
\usepackage{amsmath,amssymb,amsfonts}
\usepackage{amsfonts, amssymb, bm, bbm}
\usepackage{slashed}
\usepackage{bbm}
\usepackage{nicefrac}
\usepackage{dsfont}
\usepackage{fixmath}
\usepackage{color}
\usepackage[colorlinks=true, pdfstartview=FitV, linkcolor=red, citecolor=blue, urlcolor=black]{hyperref}
\usepackage{mathtools}

\usepackage{amssymb}
\begin{document}
\title{Evolution of holographic Fermi surface from non-minimal couplings}
\author{Wadbor Wahlang
\thanks{\emph{Email:} wadbor@iitg.ac.in}%
}                     

\institute{Department of Physics, Indian Institute of Technology, Guwahati,	Assam - 781039, India }
\date{Received: date / Revised version: date}
%
\abstract{
We study a holographic toy model by considering a probe fermion of finite charge density in an anisotropic background. By computing the fermionic spectral function numerically, we observed that the system exhibits some interesting behaviours in the nature of the Fermi surface (FS) and its evolution when tuning the controlling parameters. We introduced non-minimal interaction terms in the action for holographic fermions along with a complex scalar field but neglecting the backreaction of the fermions field on the background. Suppression in the spectral weight and deformation of FS is observed, which are reminiscent of the results seen in various condensed matter experiments in real materials.
%
     } 
%
\maketitle
\section{Introduction}
\label{intro}
The study of strongly correlated condensed matter systems using holographic approach is  one of the active field of research.  It provides the necessary tool in understanding the condensed matter systems from classical gravity theories. The earlier promising results of holographic method were those of its application in quark-gluon plasma \cite{Sadeghi:2013zma}  and in holographic superconductors.
The well-known holographic principle \cite{Maldacena:1997re,Witten:1998qj} has been applied as an alternative framework to understand various complex phenomena; such as non-linear hydrodynamics \cite{Bhattacharyya:2008jc}, non-Fermi liquid behavior \cite{Faulkner:2009wj,Cubrovic:2009ye}, transport phenomena \cite{Herzog:2007ij}, and high temperature superconductors 
\cite{Hartnoll:2008vx,Hartnoll:2008kx,Horowitz:2010gk} to point out a few.
Since then, several applications have been found out to study the emergence of Fermi and non-Fermi liquid behaviours in the fermionic systems \cite{Liu:2009dm,Lee:2008xf,Faulkner:2009am,Guarrera:2011my,Faulkner:2011tm}, and later this method was also implemented as a tools to study the conductivities and phase transition in high $T_c$ superconductor, the pseudogap phase, Mott transition  \cite{Hartnoll:2008vx,Langley:2015exa,Vanacore:2015poa,Edalati:2010ww,PhysRevD.90.126013} and  many more applications in the field of high energy and nuclear physics. 
In the recent years, new materials were discovered and they belong to a class of topological insulators, and the so-called Dirac and Weyl semimetals. These new compounds have many interesting features. Some of which have posted as a challenge to the condensed matter research community.     

Some of the results seen in condensed matter experiments on these new materials, such as those from Angle-resolved photoemission spectroscopy (ARPES) \cite{Damascelli:2003bi,2011PhRvL.106l7005K} were quite a challenge for a condensed matter theorist. The study of electronic properties in these new materials and high-$T_c$ superconductors, including the band structures, quasiparticle excitations and the Fermi surface (FS) can be done through ARPES experiment.
However, ARPES shows the presence of an unconventional pseudo-gap phase and a disconnected Fermi surface in the spectral function. This partially destructed Fermi surface is known in literature as the Fermi Arc 
\cite{1998Natur.392..157N,2006PhRvB..74v4510Y,Cremonini:2018xgj,Seo:2018hrc,PhysRevB.99.161113}. 
The appearance of Fermi arc has also been observed in the ARPES experiments in different condensed matter systems such as in the topological insulators, Dirac and Weyl semi-metals \cite{PhysRevB.99.161113,YangL,Brillaux:2020chh,Su-Yang,Su-Yang1,Noam,Peng,Mehdi,MZHazan,Binghai,PhysRevB.83.205101}. A detailed review on this topic can be found in \cite{BaiqingLv}. In all such mentioned systems, Fermi arcs appear as the surface states. It was speculated that the presence of these surface states might be caused by some unknown underlying symmetries or structures in the bulk of the material. Since these systems are strongly correlated, it is difficult to pin point the exact mechanisms that cause these unconventional electronic properties using conventional approaches. To address these problems, many have started  using AdS/CFT correspondence as a tool that allow us to study the strongly correlated systems. 
In the recent years, lots of experiments and theories were performed and modelled to shed some light in understanding of the underlying mechanism behind these strange properties of the Fermi surface from  the perspective of condensed matter physics \cite{PhysRevB.76.174501,PhysRevB.73.174501,PhysRevB.86.115118,PhysRevB.74.125110} and other approaches like the holographic method 
\cite{Hartnoll:2009sz,Herzog:2009xv,McGreevy:2009xe,Zaanen:2015oix,Hartnoll:2016apf}.   In the recent pioneering works \cite{Vanacore:2015poa,Edalati:2010ww,Edalati:2010ge,Vegh:2010fc,Benini:2010qc}, they have investigated Mott transition and the evolution of Fermi arc-like structures, taking various classical gravity background as the bulk theory. The authors in \cite{Edalati:2010ww,Edalati:2010ge}, considered a non minimal coupling of the form $p\,\bar{\psi}\,\slashed{F}\psi$, where the gauge field is coupled with the fermions field $\psi$. They observed a transfer of spectral weight between energy bands as a strength of the coupling $p$ is increased and also
	beyond certain critical value of $p$, a gap emerges in the spectral function which are the key features of Mott physics in the condensed
	matter systems. Later in \cite{Vanacore:2015poa}, by exploring the system symmetries they proposed a similar modification of the coupling mentioned earlier and the Fermi-arcs features were seen.

Motivated by the features seen in these works \cite{Vanacore:2015poa,Edalati:2010ww,Edalati:2010ge}, in this paper, we will study a holographic toy model similar to the one considered in our previous work \cite{Chakrabarti:2021qie}, where we observed an evolution of a pair of Fermi arcs. Here, we will essentially extend by modifying our previous couplings with a different one  \cite{Chakrabarti:2019gow}, where the bulk dipole coupling is modified to break the rotational and Lorentz symmetries at the boundary. Our main objective here is to further study  the evolution of the Fermi surface and their band structure in presence of the two couplings given in section \ref{sec3}.

%
%
In this work, we will consider the  Q-lattice  \cite{Donos:2013eha} as our holographic lattice background which were constructed with a complex scalar field that breaks the translational symmetry in the bulk. While in \cite{Chakrabarti:2021qie}, we have studied the
	fermionic spectral function in the Q-lattice background with couplings of the form $p\,\bar{\psi}\,\slashed{F}\psi$ and $p\,\bar{\psi}\,\slashed{F}|\phi|^2\psi$. Given the anisotropic background it  will be interesting to study the Fermi surface evolution due to Fermions bulk couplings parameters given in equation \ref{Fermionaction} below.


We will organise the discussion in this paper as follows: In section \ref{sec2}, we have a brief discussion on the important aspects of the Q-lattice background solution, then in section \ref{sec3}, we write down the Fermion's action with our coupling terms and equations of motion, leading to the spectral function $A\,(\omega,\,\vec{k})$ which is the most relevant quantity for our discussion. Thus is followed by section \ref{sec4}, where we present the numerical results with discussion on each of them. In the last section, we summarize the results and conclude with some futures directions.

\section{The Q-lattice geometry }\label{sec2}
In this section, we will briefly discuss about the Q-lattice since we had put a detail discussion in \cite{Chakrabarti:2021qie}, we will highlight only few important points here.
Here, the set up for the Q-lattice background will be a $(3+1)-$dimensional gravity theory in the bulk that could be mapped to a dual theory at the boundary in $(2+1)-$ dimensions. The full action consists of Einstein-Maxwell field and a complex scalar field given by:
\begin{align}\label{action}
	\mathcal{S}=\frac{1}{\kappa^2} \int d^{4}x\sqrt{-g}\left[\mathcal{R}+\frac{6}{L^2}-\frac{1}{4}F^2 -|\partial \phi|^2-m^2|\phi|^2\right]
\end{align}
where $\mathcal{R}$ is the Ricci scalar, Maxwell's field strength $F=dA$. Finally, $\kappa^2=16\pi G$ and $L$ are the effective reduced gravitational constant and the AdS radius respectively which we set to be unity later. As mentioned before, we consider a complex scalar field $\phi$, which will break the translational invariance of the boundary field theory.
From the above action (\ref{action}) we have the following equations of motions
\begin{align}\label{eqnom}
	&R_{\mu\nu}=\nonumber\\ &g_{\mu\nu}\left(-3+\frac{m^2}{2}|\phi|^2\right)+\partial_{(\mu}\phi\partial_{\nu)}\phi^*
	+\frac{1}{4}\left(2F_{\mu\nu}^2-\frac{1}{2}g_{\mu\nu}F^2\right)\nonumber\\
	&\nabla_{\mu}F^{\mu\nu}=0\;, \;\;\left(\nabla^2-m^2\right)\phi=0 .
\end{align}
To evaluate these equations, we shall take the following ansatz for the metric and the complex scalar field 
\begin{align}\label{ansatz}
	&ds^2=-g_{tt}(z)\;dt^2+g_{zz}(z)\;dz^2+g_{xx}(z)\;dx^2+g_{yy}(z)\;dy^2.\\
	&\text{where},\nonumber\\
	&g_{tt}(z)=\frac{U(z)}{z^2};\;g_{zz}(z)=\frac{U(z)^{-1}}{z^2};\;g_{xx}(z)=\frac{V_1(z)}{z^2};\nonumber\\
	&g_{yy}(z)=\frac{V_2(z)}{z^2},\;\;A=(1-z)\;a(z) dt,\nonumber\\
	&\;U(z)\,=\,(1-z)\;u(z),\,\,\phi=e^{i\,k_1\,x+i\,k_2\,y}\;\chi (z)\,.
\end{align}
In particular, the ansatz for the scalar field is chosen to incorporate the breaking of translational symmetry by the same scalar field in both $x$ and $y$ directions.
Here, $u,V_1,V_2,a,\chi$ are all unknown functions which depends on radial coordinate $z$, and $k_1,\,k_2$ are constants interpreted as wave numbers in the lattice. From equation \ref{eqnom} above, when combined with the ansatz, we get four second order coupled ODEs for $V_1,\,V_2,\,a,\,\chi$ and one first order for $u$. In general, for none zero scalar mass $m_{\phi}^2$, the leading behaviour of the complex scalar field near the AdS boundary ($z\longrightarrow 0$) is given by
\begin{align}
	&\chi(z)\,=\,z^{\alpha_-}\,\chi^{(1)}\,+\,z^{\alpha_+}\,\chi^{(2)}\,+ ...
\end{align} 
where, $\alpha_{\pm}=3/2\pm\sqrt{9/4+m_{\phi}^2}$\,. The leading term $\chi^{(1)}$ is associated with the source of the dual scalar operator in the boundary theory, whose dimension is $\Delta=3-\alpha_-=\alpha_+$. To this end we study in detail the case when, scalar field mass $m_{\phi}^2=0$, which corresponds to $\alpha_-=0$ and the corresponding marginal dual scalar operator in the (2+1) dimensional boundary theory. To construct  a numerical black hole solution, we solve these equations as boundary value problems with one end at the horizon and the other at the AdS boundary. At the horizon $z=1$, we use the regularity conditions by expanding the fields near the horizon, and at the AdS boundary or UV, we assume the following leading expansion,
\begin{align}
	&u=1+O(z)\,\,,\,\,V_1=1+O(z)\,\,,\,\,V_2=1+O(z)\;,\nonumber\\ &a= \mu + O(z)
\end{align} 
Once we obtained the boundary conditions, we linearize the ODEs by discretizing simultaneously both the equations of motions and the boundary conditions following \cite{Trefethen,Boyd} and then use the iterative Newton-Raphson's (NR) method following the techniques given in \cite{Andrade:2017jmt,Krikun:2018ufr}. Further, for fixed mass $m^2_{\phi}$, we found that the solutions are specified in terms of three dimensionless parameters namely, $T/\mu\,,\, k/\mu$ and  $\chi^{(1)}/\mu^{\alpha_-}$, where $ T $ is the temperature of black hole.
We showed some plots of the background solutions in our previous work \cite{Chakrabarti:2021qie} and we will use the same numerical background for analysis of the fermion spectral function later.

\section{Fermions: Action and the spectral function.}\label{sec3}

As mentioned in the introduction, we will consider the probe fermions in the anisotropic background solutions discussed above along with two non minimal couplings  and the fermionic action is given by
\begin{align}\label{Fermionaction}
	\mathcal{S}_{\psi}=\int d^4x\sqrt{-g}i\bar{\psi}\bigg(\slashed{D}-m_{\psi}-i\,p_1\,|\phi|^2\slashed{F}-i\,p_2\, |\phi|^2{\bf{\Gamma}}\slashed{F} \bigg)\psi
\end{align}
Here, $\phi$ is the same complex scalar field discussed in the lattice solution and $m_{\psi}$ is the fermion mass. Expanded expression of the symbols given in \ref{Fermionaction} are as follows:\;
\begin{align*}
	&&{\bf{\Gamma}}= \Gamma^{\underline{z}}\Gamma^{\underline{t}}(\hat{n}.\vec{\Gamma})\;,\;\;\;\;\;\;
	&\slashed{D}=e^{\mu}_c\Gamma^c\left(\partial_{\mu}+
	\frac{1}{4}\omega_{\mu}^{ab}\Gamma_{ab}-iq A_{\mu}\right)\nonumber\\
	&&\slashed{F}=\frac{1}{2}\Gamma^{ab}e^{\mu}_ae^{\nu}_bF_{\mu\nu}\;,\;\;&\vec{\Gamma}\equiv(\Gamma^{\underline{x}},\Gamma^{\underline{y}}).
\end{align*}
where,  $e^{\mu}_a ,\omega_{\mu}^{ab}$    are the vielbeins and spin connection, $q$ is the charge of the bulk fermions. 
The first  coupling with controlling parameter $p_1$ is the same to what we studied in \cite{Chakrabarti:2021qie}, however we include here to study the combine effects. While the second coupling parametrized by $p_2$ is similar the one studied in the evolution of Fermi Arc from Mott insusator \cite{Vanacore:2015poa} that breaks rotational and Lorentz symmetries of the boundary theory. We studied this type of coupling term in our previous work \cite{Chakrabarti:2019gow}, involving a real scalar field that controls the transition from circular to arc like Fermi surface controlled by the scalar condensation in the bulk at low temperature. In this paper, we will also look at the effects of these couplings on fermionic spectrum.
The Dirac equation from the above action is given by:
\begin{eqnarray}
	\label{DiracEquation1}
	\bigg(\slashed{D}-m_{\psi}-i\,p_1\,|\phi|^2\slashed{F}-i\,p_2\, |\phi|^2{\bf{\Gamma}}\slashed{F} \bigg)\psi=0.
\end{eqnarray}
In order to further simplify the Dirac equation, we choose the following gamma matrices
\begin{align}
	\label{GammaMatrices}
	&\Gamma^{\underline{z}} = \left( \begin{array}{cc}
		-\sigma^3  & 0  \\
		0 & -\sigma^3
	\end{array} \right), \;\;
	\Gamma^{\underline{t}} = \left( \begin{array}{cc}
		i \sigma^1  & 0  \\
		0 & i \sigma^1
	\end{array} \right),\nonumber\\
	&\Gamma^{\underline{x}} = \left( \begin{array}{cc}
		-\sigma^2  & 0  \\
		0 & \sigma^2
	\end{array} \right)\;\;, \;\;
	\Gamma^{\underline{y}} = \left( \begin{array}{cc}
		0  & \sigma^2  \\
		\sigma^2 & 0
	\end{array} \right)
\end{align}

\noindent Now, we can expand the Dirac equation (\ref{DiracEquation1}) together with rescaling of the field $\psi=(g_{tt}g_{xx}g_{yy})^{-\frac{1}{4}}e^{-i\omega t + ik_x x + ik_y y}\xi\,(z,\vec{k})$, where the vector $\vec{k}\equiv(-\omega,k_x,k_y)$, we have
\begin{align}
	\bigg(&\frac{1}{\sqrt{g_{zz}(z)}}\;\Gamma^{\underline{z}}\;\partial_{z}+\frac{1}{\sqrt{-g_{tt}(z)}}\;\left(\Gamma^{\underline{t}}\;(-i\omega)-iq A_{t}\right)+\nonumber\\&\frac{1}{\sqrt{g_{xx}(z)}}\;\Gamma^{\underline{x}}\;(ik_x)+\frac{1}{\sqrt{g_{yy}(z)}}\;\Gamma^{\underline{y}}\;(ik_y)-m_{\psi}\;+\nonumber\\&\frac{\partial_{z}A_t}{\sqrt{-g_{zz}(z)g_{tt}(z)}}\left(-i\,p_1\,|\phi|^2\;\Gamma^{\underline{z}\;\underline{t}}-i\,p_2\; |\phi|^2\,{\bf{\Gamma}}\;\Gamma^{\underline{z}\;\underline{t}}\right)\bigg)\,\xi(z,\textbf{k})=0
\end{align}

Now using the basis (\ref{GammaMatrices}) and by splitting the spinors $\xi=(\xi_1,\xi_2)^T$, and  $\xi_j=(\beta_j,\alpha_j)^T$, with $j$ taking values $j=1,2$, we get the following coupled radial equations  
\begin{align} 
	&
	\left(\frac{1}{\sqrt{g_{zz}}}\partial_{z}\pm m_{\psi} \right)\left( \begin{matrix} \beta_{1} \cr  \alpha_{1} \end{matrix}\right)
	\mp \frac{(\omega+ q A_{t})}{\sqrt{-g_{tt}(z)}}\left( \begin{matrix} \alpha_{1} \cr  \beta_{1} \end{matrix}\right)
	+\frac{k_x}{\sqrt{g_{xx}}}\left( \begin{matrix} \alpha_{1} \cr  \beta_{1} \end{matrix}\right)\nonumber\\&-\frac{k_y}{\sqrt{g_{xx}}}\left( \begin{matrix} \alpha_{2} \cr  \beta_{2} \end{matrix}\right)
	+\frac{\partial_{z}A_{t}}{\sqrt{-g_{zz}g_{tt}}}\;(p_1\,|\phi|^2-p_2\;|\phi|^2)\left( \begin{matrix} \alpha_{1} \cr  \beta_{1} \end{matrix}\right)
	=0
\end{align}
\begin{align} 
	&
	\left(\frac{1}{\sqrt{g_{zz}}}\partial_{z}\pm m_{\psi} \right)\left( \begin{matrix} \beta_{2} \cr  \alpha_{2} \end{matrix}\right)
	\mp \frac{(\omega+ q A_{t})}{\sqrt{-g_{tt}(z)}}\left( \begin{matrix} \alpha_{2} \cr  \beta_{2} \end{matrix}\right)
	-\frac{k_x}{\sqrt{g_{xx}}}\left( \begin{matrix} \alpha_{2} \cr  \beta_{2} \end{matrix}\right)\nonumber\\&-\frac{k_y}{\sqrt{g_{xx}}}\left( \begin{matrix} \alpha_{1} \cr  \beta_{1} \end{matrix}\right)
	+
	\frac{\partial_{z}A_{t}}{\sqrt{-g_{zz}g_{tt}}}\;(p_1\,|\phi|^2+p_2\;|\phi|^2)\left( \begin{matrix} \alpha_{2} \cr  \beta_{2} \end{matrix}\right)
	=0
\end{align}
From now on, we  follow the standard holographic procedure for extracting the Green's function. One can expand these equations near the horizon ($z=1$) to obtain the in-falling boundary condition given by
\begin{equation}
	\left( \begin{matrix} \beta_{j}(z,\vec{k}) \cr  \alpha_{j}(z,\vec{k}) \end{matrix}\right)
	\sim c_j(1-z)^{-\frac{i\omega}{4\pi T}}.
\end{equation}
Now to extract the boundary Green's function, one needs the asymptotic behaviour of the Dirac equations near the AdS boundary $(z\rightarrow 0)$ to  identify the source and the expectations values. In this case, the leading behaviour at the boundary is given by 
\begin{equation} \label{}
	\left( \begin{matrix} \beta_{j} \cr  \alpha_{j}\end{matrix}\right)
	{\approx}\; a_{j}z^{m_{\psi}}\left( \begin{matrix} 1 \cr  0 \end{matrix}\right)
	+b_{j}z^{-m_{\psi}}\left( \begin{matrix} 0 \cr 1 \end{matrix}\right).
\end{equation}
We then followed the usual prescription  used in \cite{Faulkner:2009am,Guarrera:2011my,Vanacore:2015poa} to obtain the Green's
function by extracting the coefficients $a_j$ and $b_j$ and because the couplings considered here, we have a mixing of various spinorial components, thus we use two different sets of linearly independent boundary conditions that can be written in the form $B=SA$ and expanded as
\begin{align}\label{defn}
	\left(\begin{array}{cc}\beta^I_1& \beta^{II}_1\\ \beta^I_2 & \beta^{II}_2 \end{array} \right)=  \left(\begin{array}{cc} s_{11}& s_{12}\\ s_{21} & s_{22} \end{array} \right) \left(\begin{array}{cc}\alpha^I_1& \alpha^{II}_1\\ \alpha^I_2 & \alpha^{II}_2 \end{array} \right),
\end{align}
Now the retarded Green's function is defined as   
\begin{align}\label{GreenFunction}
	G^R(\omega,k_x,k_y)=-i \left(\begin{array}{cc} s_{11}& s_{12}\\ s_{21} & s_{22} \end{array} \right)\cdot \gamma^t
\end{align}
with gamma matrices in our basis as $\gamma^t=i\sigma_1$.
Now from the retarded Green's  function we can get quantity of interest, i.e., the spectral function $A\,(\omega,k_x,k_y) $  given by
\begin{align}\label{specFunc}
	A(\omega\,,\vec{k})=\text{Im}\left[\text{Tr}\, G^R(\omega\,,\vec{k})\right].
\end{align}
where $\vec{k}\,\equiv\,(k_x,k_y)$. In the section that follows we will study the properties of $A\,(\omega\,,\vec{k}) $. The importance of this spectral function is that it directly relates to real condensed matter experiments. 
\section{Numerical Results and Discussion}\label{sec4}

In this section, we are going to look at the qualitative properties of the spectral function $A(\omega,\vec{k})$ defined in (\ref{specFunc}) above, which reflect on  the nature of Fermi surface (FS) and the dispersion spectrum by varying the coupling $p_1\,,p_2$ and also the background parameters. For our numerical purposes, we defined the Fermi level with small offset from  $\omega=0$.
In our previous work \cite{Chakrabarti:2021qie}, we have explored the affects of temperature and the source $\chi^{(1)}$ on the Fermi surface with and without couplings, also more analysis on fermions spectral function due to the Q-lattice can be found in these papers \cite{Ling2014,Iliasov:2019pav}. Here we show the results only for scenarios when  coupling parameters are non-vanishing.
\par 

\begin{figure}[t]
	\centering
	\includegraphics[width=\columnwidth]{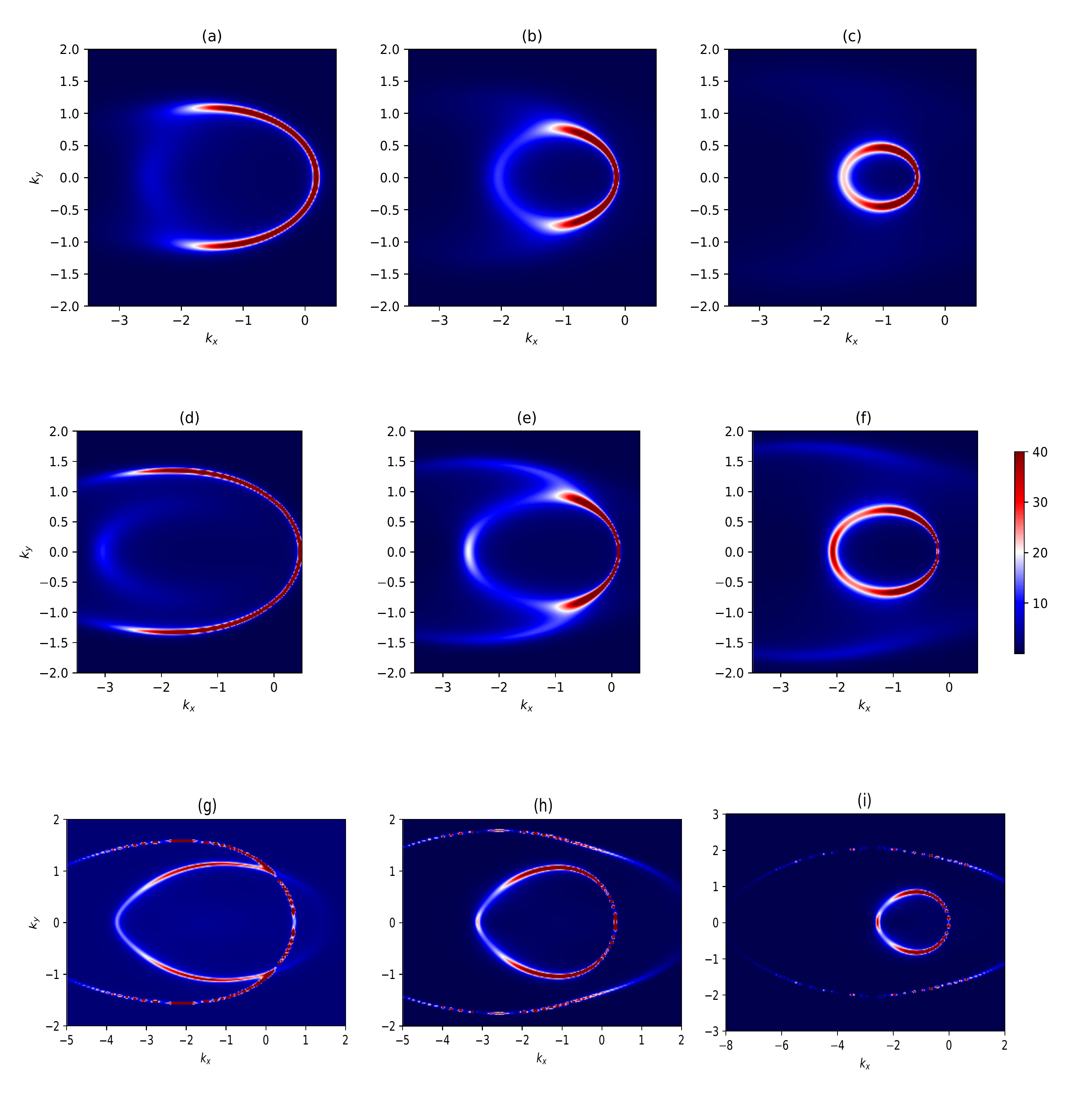}
	\caption{Density plot of $A\,(k_x,k_y)$ by varying the value of $p_1$. Here, we fixed $p_2=0.3$ and  $m_{\psi}=0\;,q=1$. Left to right: are for $p_1=-0.2\,,-0.3,\,-0.4,$ . The top, middle and bottom panel correspond to $T/\mu=0.02,\,0.009,\,0.001$ with $m_{\phi}^2=0$ and source $\chi^{(1)}=2$ respectively.}
	\label{P1P2combined}
\end{figure}

As discussed above the complex scalar field can break translational invariance in both $x$ and $y$ directions, depending on $k_1$ and $k_2$. Let us first study for $k_2=0$ and consider only $k_1$.
Fixing the background scalar mass $m_{\phi}^2=0$, then the leading constant $\chi^{(1)}$ which represents the source of dual boundary operator of the bulk complex scalar field can be varied and we have seen in \cite{Chakrabarti:2021qie} and references therein, as we increase the strength of the source, the weight of spectral function reduces along $k_x$ where translational symmetry is broken. Since the spectrum of our first couplings control by $p_1$ is known from our previous discussion in \cite{Chakrabarti:2021qie}, we directly show the combined effects of $p_1$ and $p_2$ in
\figurename{ \ref{P1P2combined}}.
\begin{figure}[t]
	\centering
	\includegraphics[width=\columnwidth]{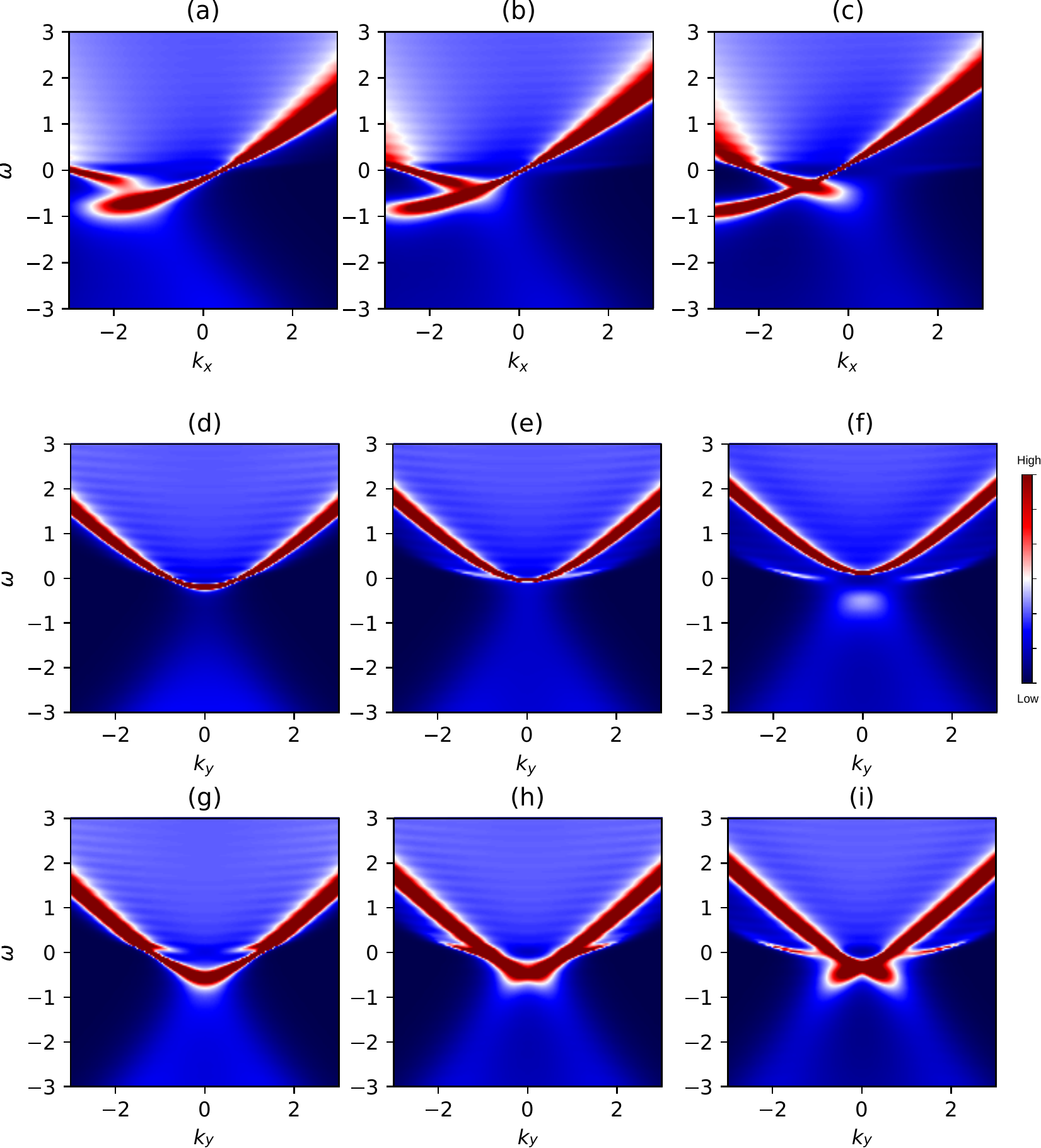}
	\caption{Energy-momentum dispersion  with $p_1$,\, $p_2$ set to non-zero and  $m_{\psi}=0,\;q=1$,\, $T/\mu=0.009$,\;$m_{\phi}^2=0$,\; $k_1/\mu=0.8$,and\,$k_2/\mu=0$. Top and middle panel are along $\omega-k_x\;(k_y=0)$ and  $\omega-k_y\;(k_x=0)$ plane, while the bottom panel is along $\omega-k_y\;(k_x=-1)$. From left to right the parameters [\,$(p_2\,,p_1)\,]\,=\,[\,(0.3,-0.2),(0.3,-0.3),(0.3,-0.4)]$ respectively.}
	\label{P3P4combinedomk}
\end{figure}
One can immediately notice in \figurename{ \ref{P1P2combined}} that for low temperature (bottom panel) the combine effects yields two arcs like spectrum, and the peak of FS is more sharp in comparison to that of higher temperature (top panel) where the inner arc vanishes. Further, the stretched Fermi surface that appeared in the spectral function because of the anisotropic background for large $\chi^{(1)}$ values, now its amplitude is suppressed and more prominent along negative $k_x-$direction specially at low temperature, due to the second coupling. However, from previous analysis at lower temperature, the effects of $p_1$ induces a second Fermi surface that is partially destroyed along $x-$direction which is the signature of second coupling. Whereas, along $k_y$ direction, the spectral function indicates the existence of sharp Fermi surface with a large density of states. The sharp peak in the spectral function, essentially indicate the presence of stable quasi-particle with longer lifetime.   
Similarly, from the energy momentum dispersion shown in \figurename{ \ref{P3P4combinedomk}}, we see the presence of gapless spectra which is anisotropic in nature and the surface state induces by $p_2$ connecting the lower and upper band.\par
From the emergence of these interesting behaviour in the spectral function from our couplings $p_1$ and $p_2$,  firstly we can think of the effects of the source $\chi^{(1)}$ as a doping parameter in the pseudogap region and secondly, for both the couplings, we know from the work in \cite{Vanacore:2015poa,Edalati:2010ww}, the appearance of these Fermi arcs can be understood in the context of pole-zero duality \cite{PhysRevD.90.126013}. In terms of the two diagonal entries of the boundary fermions Green's function $G_{ij}(\omega,\vec{k})$, the FS appears at two poles which exist at $G_{11}(\omega=0,|k_f|)$ and $G_{22}(\omega=0,-|k_f|)$. Now the presence of $p_2$ coupling, whenever the pole/zero of $G_{11}$ coincides with zero/pole of $G_{22}$, we get a reduced spectral weight along $-k_x$ or $+k_x$ depending on the sign of $p_2$ that appears as Fermi arcs. The presence of these pairs of FS in the spectral function indicates that the scalar field plays a crucial role in not mixing the individual spectra of $p_1$ and $p_2$. Hence we eventually have two Fermi surfaces on changing $p_1,p_2$. Notice that in our plots, we choose the mass of scalar field to be $m_{\phi} =0$, because for non-zero masses, the way our scalar field manifest itself in the coupling terms, its effects will be highly suppressed. However, we plotted for non-zero fermion mass in \figurename{ \ref{nonzeromassvaryp3p4}. Again here for same charge $q$ and background parameter we see a similar spectrum to that in the case of zero mass shown above.\par 
	So far we have consider only $k_1$, now let us turn on $k_2$. As shown in our previous work \cite{Chakrabarti:2021qie}, depending on whether $k_1>k_2$ or $k_1<k_2$, the fermionic spectrum is gapped along $k_x$ or $k_y$. We consider the case when $k_2>k_1$ and the plots are shown in \figurename{ \ref{varyP2varyL}. In the top panel we fixed $p_1=1$ and vary boundary source $\chi^{(1)}$ of the bulk scalar field $\phi$. As the source $\chi^{(1)}$ is increased, in addition to being stretch from circular shape to an elliptical shape, we see an emergence of a second FS of smaller peaks. Some similarity is seen when we vary $p_2$ instead and fixing $\chi^{(1)}=2$. Here also we observed a new kind of spectrum which is different from that observed in \cite{Vanacore:2015poa,Chakrabarti:2019gow} with similar couplings. 
	It was earlier shown in \cite{Gubser:2012yb} the presence of multiple FS in holographic model with no Fermi arcs and then recently in \cite{Cremonini:2019fzz,Balm:2019dxk}, the Fermi features were seen triggered by the effective mass term and the lattice effects in their holographic models. It is to be noted here that though both our studies and theirs are somewhat similar in terms of the features in suppression of the spectral function and the Fermi surfaces, our couplings and the one considered in \cite{Cremonini:2019fzz} should correspond to a different boundary field theory with a different kind of underlying mechanism. Moreover, the evolution of these Fermi arcs without including superconductivity as discussed in \cite{Vegh:2010fc,Benini:2010qc}, it could provide an interesting framework for studying the Fermi arcs in the pseudogap phase and  cuprates. 
		
		Up till now, we have discussed only the spectra produced by the two couplings. Let us try to relate these results to real condensed matter systems. Firstly, the emergence of these pair of Fermi surfaces are mostly our interesting features out of our numerics as these kind of spectra have been observed in the ARPES experiments measuring the spectral function in real condensed matter systems such as those of Topological insulators (TI), Dirac and Weyl semi-metals \cite{PhysRevB.99.161113,YangL,MZHazan}. Also, for superconductivity, these types of interactions terms provides us a holographic model in the pseudogap region in the phase diagram where these Fermi arcs are seen. They are connected through the suppression in the weight spectral function. Similar to our observation, the presence of these kind of arcs in pairs, was also discussed in a theoretical model in \cite{Kun-Yang}. While in the case of Dirac and Weyl semi-metals, Fermi arcs emerges out as surface states which connect the two Dirac points in three-dimensional bulk crystals. From the energy band dispersion shown in \figurename{ \ref{P3P4combinedomk}} in our models,  we can see the presence of non-trivial gapless surface states that were also observed in experiments. Details discussion on these topics can be found in these articles  \cite{Su-Yang,Su-Yang1,Noam,Peng,MZHazan} and references therein. Since our holographic system at the boundary is ($2+1$) dimensions, we can think of these pairs of Fermi arcs as the surface states of a ($3+1$) dimensional bulk material. Thus the observations from our holographic models are reminiscent to those of real materials in terms of the band spectrum and surface states known as Fermi arcs. There have been other holographic models also, such as those in \cite{Rodgers:2021azg,Landsteiner:2019kxb,Oh:2020cym}, which attempted to address these interesting and rich phenomena observed in topological insulators and semi-metals.
		\begin{figure}[t]
			\centering
			\includegraphics[width=\columnwidth]{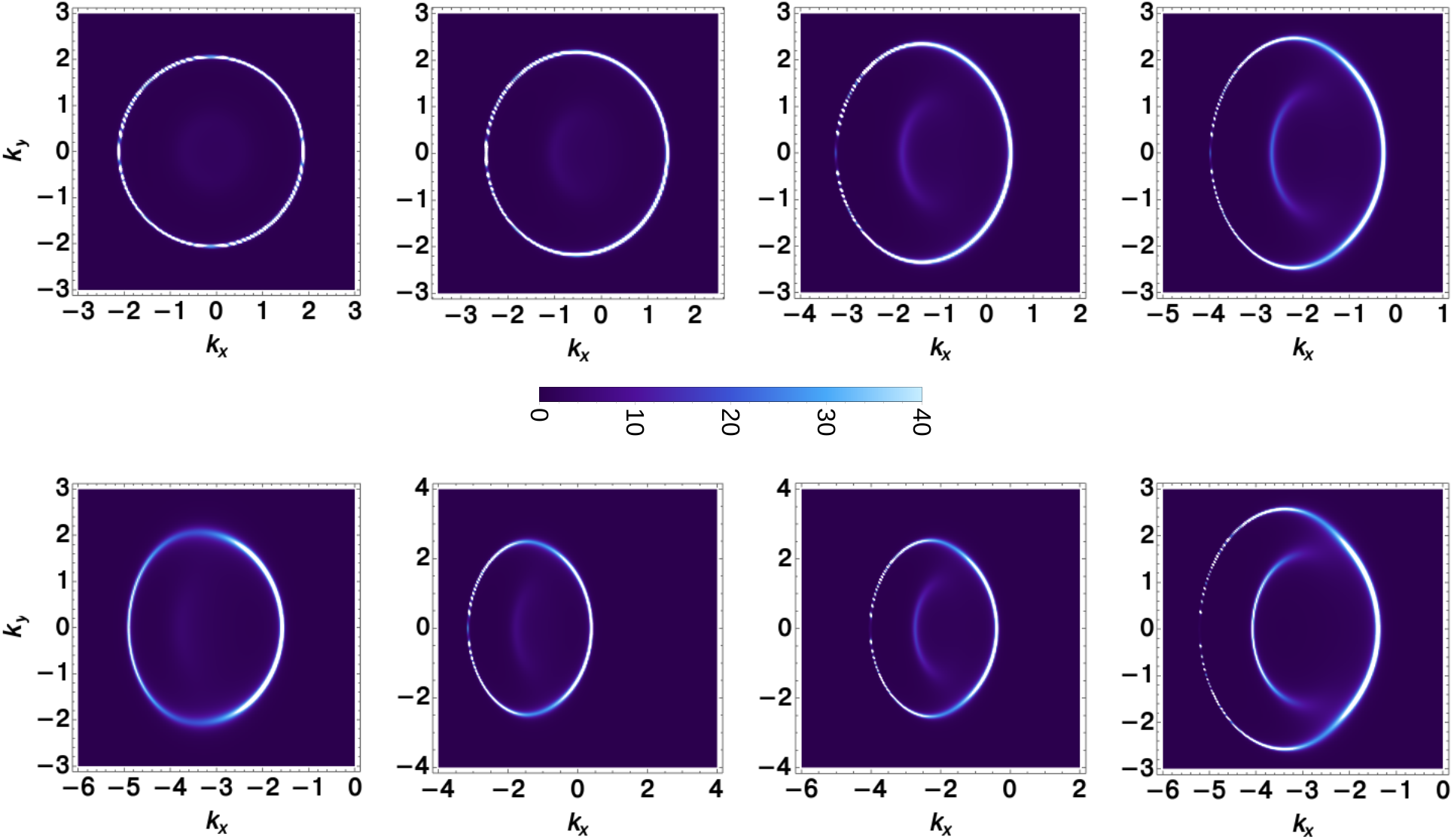}
			\caption{Here we plotted the spectral density with fermion mass $m_{\psi}=0$ and charge $q=1$. In the top panel, we fixed $T/\mu=\,0.009$,\,coupling $p_2=1.$ and vary source  $\chi^{(1)}=(\,0.5,\,1.0,\,1.5,\,1.8\,)$  with $k_1/\mu=0.2$ , $k_2/\mu=0.8$, and $m_{\phi}^2=0$ respectively. In the bottom panel, we vary the coupling $p_2$\; taking values $p_2=0.1,\,0.5,\,0.8,\,1.2$ while we fixed $\chi^{(1)}=2$.}
			\label{varyP2varyL}
		\end{figure}
		\begin{figure}[htbp]
			\centering
			\includegraphics[width=\columnwidth]{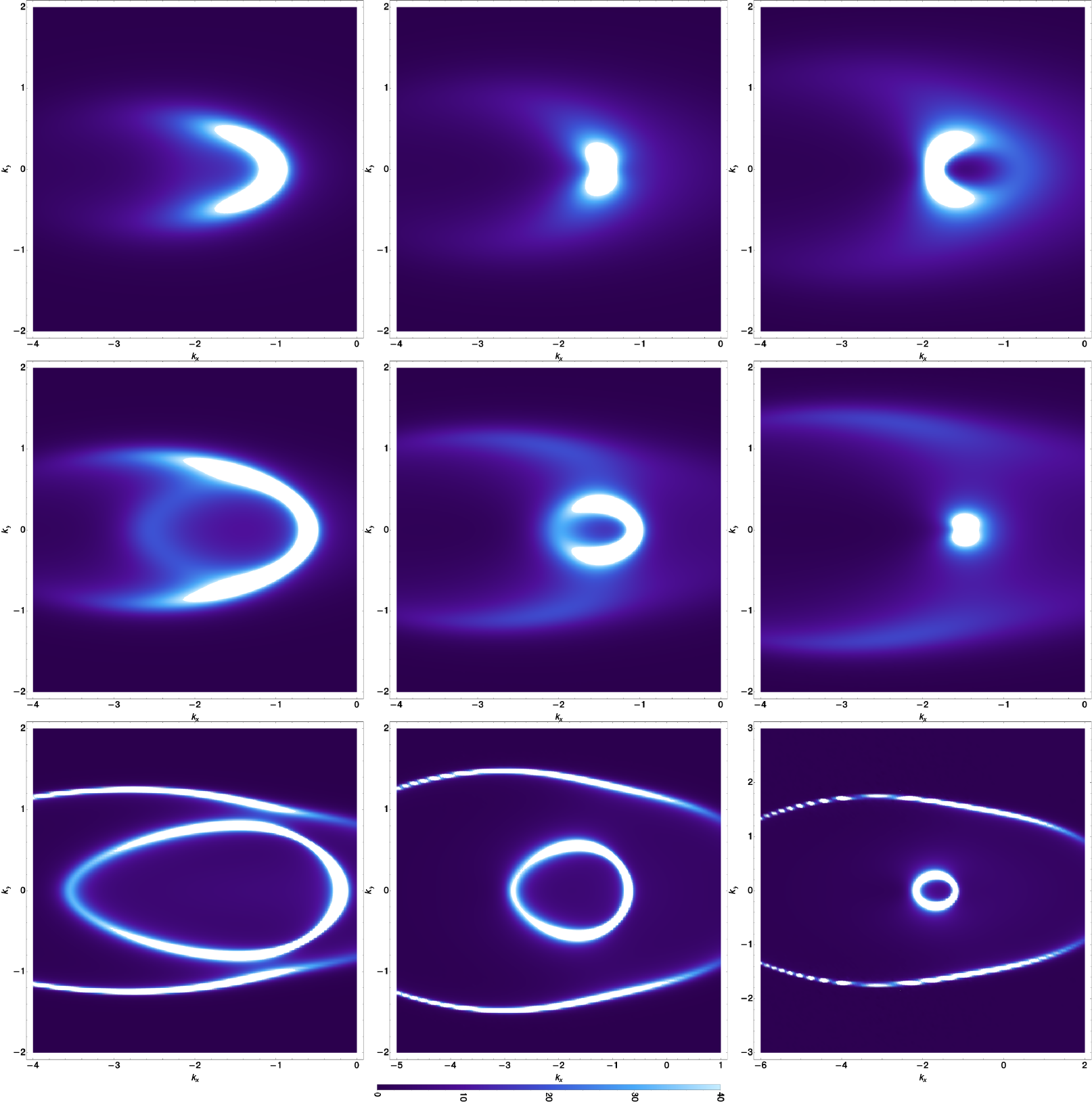}
			\caption{Here we plotted the spectral density for non-zero fermion mass $m_{\psi}=1/4$ and charge $q=1$ with the background parameters  $T/\mu=0.02,\,0.009,\,0.001$ (top to bottom), $k_1/\mu=0.8$,\,$k_2/\mu=0$, $m_{\phi}^2=0$ and $\chi^{(1)}=2$ respectively. We fixed the coupling parameters  $p_2=0.3$ and vary $p_1=-0.2,\,-0.3,\,-0.4$.}
			\label{nonzeromassvaryp3p4}
		\end{figure}

\section{Conclusion}\label{sec5}
In this paper, we have studied the fermion spectral function $A(\omega,\vec{k})$ in anisotropic Q-lattice background where, the probe fermion is coupled with two non-minimal dipole-type couplings. We discussed some of the interesting features produced by the couplings. Our observation of the spectral function from the numerics of our holographic models seems to encode the unconventional properties that were observed in real condensed matter systems, such as the gapless spectra in the energy-momentum dispersion and the presence of Fermi arcs, also the appearance of doubled FS connecting the two nodes seen in Dirac -Weyl semimetals and topological insulators. Though we are not in anyway claiming that our models to represent the exact match with actual condensed matter systems, however we hope that in future, this kind of holographic approach  might give us some hints in understanding the strongly coupled systems.

\vskip 2mm
\noindent
\section*{Acknowledgements}
\vskip 2mm
\noindent
The author would like to thank Dr S. Chakrabarti and Dr D. Maity for their illuminating discussions and comments on the draft of this work.
%

\end{document}